\documentclass[letterpaper, 10 pt, conference]{ieeeconf}  

\IEEEoverridecommandlockouts                              
\newcommand{\beq}{\begin{equation}}
\newcommand{\eeq}{\end{equation}}
 
  \let\labelindent\relax
\usepackage{enumitem}
\usepackage{balance}

\newcommand{\bag}{\begin{aligned}}
\newcommand{\eag}{\end{aligned}}
\usepackage{float}
\newcommand{\sign}{\operatorname{sign}}
\usepackage{amssymb}
\usepackage{graphicx}

\usepackage{amsmath, amssymb}
\usepackage{balance}

\newtheorem{theorem}{Theorem}[section]

\newtheorem{remark}{Remark}

\usepackage{color}
\usepackage{bm}
\usepackage{balance}

\usepackage{url} 

 \newenvironment{IEEEkeywords}
{\textbf{Keywords---}} 
{}

\title{\LARGE \bf Finite-Time Stabilization of a Class of Nonlinear Systems in Hilbert Space}

\author{Kamal Fenza, Moussa Labbadi and Mohamed Ouzahra
\thanks{*The second author gratefully acknowledges the support of the CNRS through the project ROC--MAS LA--SE D\'EVELOP RELAT$^{\circ}$ INTERNATIO.}
\thanks{K. Fenza and M. Ouzahra are with the  Laboratory of  Mathematics and Applications to Engineering Sciences,  Sidi Mohamed Ben Abdellah University, Fes, Morocco (e-mail: \textsl{kamal.fenza@usmba.ac.ma}, \textsl{mohamed.ouzahra@usmba.ac.ma})}
\thanks{M. Labbadi is with the Aix-Marseille University, LIS UMR CNRS 7020, 13013 Marseille, France (e-mail: \textsl{moussa.labbadi@lis-lab.fr}). 
}
}

\begin{document}

\maketitle
\thispagestyle{empty}
\pagestyle{empty}

\begin{abstract}
This paper deals with the finite-time stabilization of a class of nonlinear infinite-dimensional systems. First, we consider a bounded matched perturbation in its linear form. It is shown that by using a set-valued function, both the convergence objective (finite-time) and the rejection of perturbations are achieved. Second, we consider a class of nonlinear systems and design a feedback control that ensures the closed-loop system is finite-time stable. All proofs presented in this paper regarding convergence are based on Lyapunov theory. The existence of solutions to the closed-loop system and its well-posedness are established using maximal monotone theory. To illustrate the applicability of the theoretical results, a heat equation is considered as an application of the main results.
\end{abstract}

\section{Introduction}
This paper investigates the stabilization problem for a class of linear/nonliner infinite-dimensional systems characterized by:
 bounded linear control operators; matched disturbances (cf. \cite{curtain2020, bensoussan2007, balogoun2025sliding} for comprehensive treatments of such systems). Our specific contribution focuses on the design of:
 finite-time  control schemes in the presence of matched perturbations based on set-valued sign function and nonlinear feedback~{\color{blue}\cite{utkin2019control}}.

First, we focus on the finite-time stabilization of \eqref{systSvvb_simple}, which involves finding a feedback control and a finite settling time $T$ such that all trajectories of \eqref{systSvvb} converge to zero within this time and remain at zero afterward.
Finite-time stability has important applications in control theory, robotics, and the stabilization of underwater vehicles, satellites, and automobiles, where strict time constraints are imposed \cite{efimov2021finite}.
 In the \text{finite-dimensional case}, this problem was studied for continuous autonomous systems by \cite{bhat2000finite}.
 For \text{linear systems}, \cite{6666666} addressed it using a time-varying feedback (prescribed-time convergence).
  The problem of \text{finite-time partial stability} was investigated in \cite{25} using continuous or discontinuous state feedback laws.

In the context of infinite-dimensional systems~{\color{blue} \cite{orlov2002discontinuous, cortes2006finite,coron2017null,efimov2021finite,coron2021boundary,nguyen2024rapid,polyakov2016finite,han2024input, 10, 15, 17} }extended the notion of homogeneity to infinite-dimensional systems. They developed a linear stabilizing feedback and a Lyapunov function for homogeneous dilation to obtain finite-time stability results, which were then applied to the wave and heat equations.
Additionally, the author in \cite{2333} proved finite-time stability for the \text{bilinear reaction-diffusion equation} at a settling time that depends on the initial state (in the absence of disturbances).
   It was shown that finite-time stability can be achieved at any prescribed time using a time-varying feedback.
     The finite-time internal stabilization of a linear $1D$ transport equation was considered in \cite{sogore2020}. 

A powerful tool for achieving
finite-time stabilization and rejecting matched perturbations is sliding mode control (SMC), particularly in the control of dynamical systems \cite{sogore2020, sob, balogoun2023, pilloni2024sliding,pisano2012boundary,orlov2004robust,pisano2011tracking,orlov2013boundary,orlov1995sliding}. It has been extensively studied in the context of infinite-dimensional systems, where representation and state-space methods provide a solid theoretical foundation for analysis and controller design \cite{bensoussan2007, curtain2020}. In hyperbolic and parabolic PDE control, SMC strategies, including super-twisting algorithms, have proven effective in ensuring robustness and finite-time convergence despite system uncertainties \cite{coron2021, guo2012, guo2013, balogoun2023}.  These approaches  make use of  Lyapunov-based techniques to construct strict stability proofs, as demonstrated in the development of super-twisting controllers \cite{moreno2012}.
Furthermore, boundary sliding mode control methods offer robust solutions for handling disturbances in heat and wave equations, where conventional control techniques have difficulty \cite{pisano2012, guo2012, wang2015}. For instance, second-order sliding mode controllers have been successfully applied to heat processes with unbounded perturbations, ensuring strong disturbance rejection capabilities \cite{pisano2012}. Similarly, the combination of SMC with active disturbance rejection control (ADRC) has led to significant improvements in the stabilization of wave and Euler-Bernoulli beam equations under boundary disturbances \cite{guo2012, guo2013}. The application of SMC to cascaded PDE–ODE systems has further demonstrated its flexibility in managing complex multi-scale dynamics, providing efficient stabilization mechanisms even under severe input constraints \cite{wang2015}.
By integrating SMC with these advanced techniques, researchers continue to expand its applicability to various classes of PDEs, ensuring both theoretical rigor and practical feasibility in real-world control systems.


In this work, we develop a multi-valued feedback approach for the global finite-time stabilization of a linear system subjected to perturbations. We show that the support system is finite-time stable and provide an estimation of its settling time. Furthermore, we consider a class of nonlinear systems for which we propose a nonlinear feedback control that ensures the closed-loop system is finite-time stable. As an application, we illustrate our approach using a heat equation. The main contributions of this study are as follows:
\begin{itemize}
    \item 

    Existence, uniqueness, and regularity of solutions: We rigorously establish the existence, uniqueness, and regularity of the global weak solution of the closed-loop system. This analysis is carried out using nonsmooth convex analysis and nonlinear semigroup theory, leveraging maximal monotone operators to ensure well-posedness.

      \item   Finite-time stabilization via control design: We design a systematic control strategy that guarantees finite-time stabilization of the system. The approach relies on set-valued feedback laws to handle perturbations effectively while ensuring robustness and stability.

       \item  Extension to nonlinear systems: We extended our study to account for additional nonlinearities, introducing new analytical challenges. To address these complexities, we refine the control design and stability analysis, ensuring that the proposed method remains effective under broader conditions.
    \end{itemize}
\subsection*{Notations:}
\begin{itemize}
    \item $\mathbb{R}$ denotes the field of real numbers, with $\mathbb{R}^+ = [0, +\infty)$,  $\mathbb{R}^{-} = (-\infty,0]$  {\color{blue} and \(\mathbb{R}^* = \mathbb{R} \setminus \{0\}\) denotes the set of nonzero real numbers.}
    \item $\mathcal{H}$ is a real Hilbert space endowed with the inner product $\langle \cdot, \cdot \rangle$ and its corresponding norm $\|\cdot\|$. 
    \item Let $C(\mathbb{R}^{+}, \mathcal{H})$ denote the space of all continuous functions defined on $\mathbb{R}^+$ and taking values in $\mathcal{H}$.
    \item $I$ represents the identity operator on $\mathcal{H}$. 
    \item $B^{*}$ is the adjoint of the operator $B$.
    \item Let \( A: D(A) \subset \mathcal{H} \to \mathcal{H} \) be a (possibly) unbounded operator, where \( D(A) \) denotes its domain.
    \item $\mathcal{H}^p(\Omega, \mathbb{R})$ denotes the Sobolev space of functions $\Omega \to \mathbb{R}$, where $\Omega \subset \mathbb{R}^n$ is an open connected set with a smooth boundary.
    \item $L^\infty(\Omega)$ is the space of essentially bounded, Bochner-measurable functions with the norm $\|f\|_\infty := \operatorname{ess\,sup}_{s \in \Omega} \|f(s)\|$.
    \item $C_c^\infty(\Omega)$ is the set of infinitely smooth functions $\mathbb{R}^n \to \mathbb{R}$ with compact support in $\Omega$.
    \item $\mathcal{H}_0^p(\Omega, \mathbb{R})$ is the closure of $C_c^\infty(\Omega)$ in the norm of $\mathcal{H}^p(\Omega, \mathbb{R})$.
   \item  \textcolor{blue}{ $1_E$ is the characteristic function of the set $E$}.
\end{itemize}

\section{Problem Statement}

Consider the following nonlinear  system:
\begin{eqnarray}\label{systSvvb}
    \begin{cases}
        \frac{d}{dt} y(t) = A y(t) + f(y(t))y(t) + Bu(y(t)), & t > 0, \\
        y(0) = y_{0},
    \end{cases}
\end{eqnarray}
where $y(\cdot)$ is the state, and the control input $u$ is a $\mathcal{U}$-valued function, with $\mathcal{U}$ being a real Hilbert space. The state space is a Hilbert space $\mathcal{H}$, the unbounded linear operator $A: D(A) \rightarrow \mathcal{H}$ is the infinitesimal generator of a strongly continuous semigroup $S(t)$ on $\mathcal{H}$, and $B: \mathcal{U} \rightarrow \mathcal{H}$ is a linear bounded operator. Furthermore, $f: \mathcal{H} \rightarrow \mathbb{R}^{-}$ is a scalar function  and \textcolor{blue}{let us define   the operator $F: \mathcal{H}\rightarrow \mathcal{H} $ by $F(y) = - f(y)y$}.

In this paper, we focus on the problem of \text{finite-time stabilization} of \eqref{systSvvb}, which consists of finding a feedback control and a finite (settling) time $T$ such that all trajectories of \eqref{systSvvb} converge to zero within this time and remain at zero afterward.

We make the following assumptions:
\begin{itemize}
    \item[$(\mathbf{H}_{1})$] The map $F$ can be decomposed as $F = F_{1} + F_{2}$, where $F_{1}$ is maximal monotone on $\mathcal{H}$ and $F_{2}$ is a Lipschitz operator on $\mathcal{H}$.
    Specifically, \( F_2 \) is a Lipschitz operator if the following condition holds:
\[
\|F_2(s_1) - F_2(s_2)\| \leq \kappa \|s_1 - s_2\|, \quad \forall (s_1, s_2) \in \mathcal{H}^2,
\]
where \( \kappa \) is a positive constant.
    \item[$(\mathbf{H}_{2})$] There exists $\beta > 0$ such that $\|B^{*}y\|^{2}_{\mathcal{U}} \geq \beta \|y\|^2$ for all $y \in \mathcal{H}$.
\end{itemize}

\begin{remark}
The assumption $(\mathbf{H}_{2})$ implies that the operator $BB^{*}$ is coercive. This assumption has been considered in \cite{sogore2020} and \cite{sob}.
\end{remark}
The system \eqref{systSvvb} in the presence of nonlinearity is well-studied in the literature using the theory of evolution semigroups \cite{pazy2012semigroups,engel2000one}. Recently, the work in \cite{polyakov2021input} considered a class of nonlinear evolution equations that are locally Lipschitz and addressed their ISS using homogeneous infinite-dimensional systems.

Before analyzing the complex dynamics of the system \eqref{systSvvb}, we first examine a fundamental case study---the finite-time feedback stabilization problem for semilinear parabolic systems---which serves as both a conceptual foundation and a technical prototype for our approach.

\begin{eqnarray}\label{systSvvb_simple}
    \begin{cases}
        \frac{d}{dt} y(t,x)  = \Delta y(t,x) +  a(x) u(y(t,x))  + f(t), \\
       \quad \quad \quad (t,x) \in \mathbb{R}^{+} \times \Omega, \\
        y(t,x) = 0, \quad (t,x) \in \mathbb{R}^{+} \times \partial \Omega, \\
        y(0,x) = y_{0}, \quad x \in \Omega,
    \end{cases}
\end{eqnarray}
where $\Omega \subset \mathbb{R}^{n}$ ($n \geq 1$) is an open bounded domain with smooth boundary $\partial \Omega$, and $y(t) = y(t, \cdot) \in \mathcal{H} = L^{2}(\Omega)$ is the corresponding state of \eqref{systSvvb_simple}. Here, $u$ is the control and  \textcolor{blue}{$\Delta y(x) = \sum\limits_{i=1}^n \frac{\partial^2 y}{\partial x_i^2}(x)$, for all $y\in \mathcal{H}_{0}^{1}(\Omega ) \cap \mathcal{H}^2(\Omega )$. }  The real-valued function $a: \Omega \rightarrow \mathbb{R}^{*}$ is  such that $a(x)\ne 0$, for all $x\in \Omega $) and $f$ is the perturbation term satisfying $f \in L^{\infty}(\mathbb{R}^{+})$.\\
Semilinear reaction--diffusion systems like \eqref{systSvvb_simple} can describe various models, such as nuclear chain reactions, tumor growth, and biomedical processes (see \cite{K} and \cite{Kh}).

\section{Main Results}
\subsection{A multivalued feedback approach to finite-time stabilization of perturbed system case \eqref{systSvvb_simple}}
We consider the following feedback control:
 \begin{equation}\label{control}
	u(y(t,x))=- \frac{\rho}{  a(x) }~  \sign(y(t,x)),
\end{equation}
where  $\rho >0$ is the gain control 
 and the set-valued map is defined as:
\begin{equation}
	\operatorname{sign} (y) \begin{cases}  \in \{\frac{y}{|y|}\} & \text { if}~  y ~ \neq 0, \\ = [-1,1] & \text { if } y=0.\end{cases}
\end{equation}
For our purpose, \textcolor{blue}{combining {\color{blue}(\ref{systSvvb_simple})} and {\color{blue}(\ref{control})} leads to the following differential inclusion}:
\begin{eqnarray}\label{KAMAKkbbb}
	\begin{cases}
		\frac{d }{d t}y(t,x) +\mathcal{B} (y(t,x))-f(t) \ni 0, ~~ &   (t,x)\in\mathbb{R}^{+}\times \Omega, \\
		y(t,x)=0, ~~ &  t\in \mathbb{R}^{+} \times \partial \Omega, \\
		y(0, x)=y_{0}, ~~ & x \in \Omega,
	\end{cases}
\end{eqnarray} 
with  $\mathcal{B}: \mathcal{D}(\mathcal{B}) \rightarrow P(L^{2}(\Omega))$
\[
\begin{aligned}
    \mathcal{B} y &= \left\{ z \in L^{2}(\Omega) \;\middle|\; 
    \begin{aligned}
        &z = -\Delta y + \rho w, \\
        &w(x) \in \operatorname{sign}(y(x)), \quad \text{a.e. } x \in \Omega
    \end{aligned} \right\}, \\
    \mathcal{D}(\mathcal{B}) &= \left\{ y \in \mathcal{H}_{0}^{1}(\Omega) \cap \mathcal{H}^2(\Omega) \;\middle|\; 
    \begin{aligned}
        \exists w \in L^{2}(\Omega), \\
        w(x) \in \operatorname{sign}(y(x)), \\
        \quad \text{a.e. } x \in \Omega
    \end{aligned} \right\}.
\end{aligned}
\]

In the following result  we give a representation of the solution of the differential inclusion system {\color{blue} (\ref{KAMAKkbbb})} that is based on nonlinear semigroup theory.
\begin{theorem}\label{th1}
Let $f\in L^{\infty}(\mathbb{R}^{+}; \hspace*{0.1cm}\mathcal{H})$. The operator $\mathcal{B}$ with domain $D(\mathcal{B})$ generates a nonlinear semigroup on $\mathcal{H}$ denoted by $\exp(t\mathcal{A}) $. Then, for all $y_{0}\in L^{2}(\Omega)$,  the  system {\color{blue} (\ref{KAMAKkbbb})} has a unique strong solution $y\in C([0,+\infty), \mathcal{H})$.
\end{theorem} 
\begin{proof}
 The operator $\mathcal{B}$ is maximal monotone as the subgradient of a  lower semicontinuous proper convex function on $\mathcal{H}$ (see \cite{barbu1993analysis}, Proposition 2.20).  Then,  $\mathcal{B}$ with domain $D(\mathcal{B})$  generates a  nonlinear contraction $C_{0}$-semigroup (see \cite{brezis1973operateurs}, p. 65). \\
As a consequence of the Proposition 4.1  in \cite{barbu1993analysis}, the feedback system {\color{blue} (\ref{KAMAKkbbb})} admits a unique  strong solution  $y\in C([0,+\infty), \mathcal{H})$.
\end{proof}
We now state our main result concerning the global finite-time stabilization of (\ref{KAMAKkbbb}) toward the origin equilibrium, requiring that $f(t) \in [-\rho, \rho]$, for all $t \geq 0$.
\begin{theorem}\label{th2}
	Let $f\in L^{\infty}(\mathbb{R}^{+})$.  Assume that  the gain control is  such that $\rho > \left\|f\right\|_{ L^{\infty}(\mathbb{R}^{+}; \hspace*{0.1cm}\mathcal{H})}$.
    \begin{enumerate}
        \item 
   If  $y_{0}\in L^{\infty}(\Omega)$, then the system {\color{blue} (\ref{KAMAKkbbb})} is globally finite-time stable with a settling time $T\leq \frac{\left\|y_{0}\right\|_{L^{\infty}(\Omega)}}{\rho-\left\|f\right\|_{L^{\infty}(\mathbb{R}^{+})}}$.\\
	 \item  If $y_{0}\in L^{2}(\Omega)$, then the system (\ref{systSvvb_simple}) is  globally finite-time stable under the following control:
	 \begin{equation}
	u(t,x)= 
	\begin{cases}
		0 & \text{if } 0 \leq t \leq \theta, \\
		- \dfrac{\rho}{a(x)}\, \sign\big( y(t,x) \big) & \text{if } t > \theta.
	\end{cases}
	\label{eq:control_lawy}
\end{equation}
	 The settling time can be estimated as 
	 $$
	T\leq \theta+\frac{\left\|y(\theta)\right\|_{L^{\infty}(\Omega)}}{\rho-\left\|f\right\|_{L^{\infty}(\mathbb{R}^{+})}}, \textit{and }  \theta>0.
	$$    
    \end{enumerate}
\end{theorem} 
\begin{proof} \emph{Case 1:}  Suppose that $y_{0}\in L^{\infty}(\Omega)$ and let us consider the function defined by  $\Upsilon (t, x)=\left\|y_{0}\right\|_{L^{\infty}(\Omega)}-t(\rho -\left\|f\right\|_{L^{\infty}(\mathbb{R}^{+})})$, for all  $(x,t)\in \Omega \times(0, \gamma^{-1}\left\|y_{0}\right\|_{L^{\infty}(\Omega)})=Q_{0}$ with $\gamma=(\rho  -\left\|f\right\|_{L^{\infty}(\mathbb{R}^{+})})$.\\ 
Then, we have that
\begin{IEEEeqnarray}{c}\label{KAMAKkbbbbv}
\left\{
\begin{aligned}
&\frac{\partial \Upsilon}{\partial t} - \Delta \Upsilon + \rho \operatorname{sign}(\Upsilon) 
- \left\|f\right\|_{L^{\infty}(\mathbb{R}^{+})} = 0, 
\quad \text{in } Q_{0}, \\ 
&\Upsilon(0) = \left\|y_{0}\right\|_{L^{\infty}(\Omega)},
\quad  \text{in } \Omega, \\
&\Upsilon \geq 0, 
\quad  \text{in } \partial\Omega \times (0, \gamma^{-1}\left\|y_{0}\right\|_{L^{\infty}(\Omega)}).
\end{aligned}
\right.
\end{IEEEeqnarray}
 Subtracting {\color{blue} (\ref{KAMAKkbbb})} and {\color{blue} (\ref{KAMAKkbbbbv})} and multiplying the resulting equation by $(y-\Upsilon)^{+}=\max(y-\Upsilon, 0)$, and integrating on $\Omega$ we get 
\begin{equation}\label{eq:energy_estimate}
\begin{aligned}
\frac{1}{2}\frac{d}{dt} \left\|(y-\Upsilon)^{+}(t)\right\|_{L^{2}(\Omega)}^{2} 
+ \int_{\Omega} \left( \nabla(y-\Upsilon)^{+} \right)^{2} dx \\
+ \int_{\Omega} \rho\bigl(\operatorname{sign}(y) - \operatorname{sign}(\Upsilon)\bigr)(y-\Upsilon)^{+} dx \leq  0, 
\quad \text{in } Q_{0}.
\end{aligned}
\end{equation}
\textcolor{blue}{   Using the fact that $\Upsilon(t,x) > 0$ for all $x \in \Omega$, we obtain
\[
\int_{\Omega}
\rho \big( \operatorname{sign}(y) - \operatorname{sign}(\Upsilon) \big)
\, (y - \Upsilon)^{+} \, dx = 0,
\quad \text{in } Q_{0}.
\]
This, together with \((8)\), yields
\[
\frac{d}{dt}
\big\| (y - \Upsilon)^{+}(t) \big\|_{L^{2}(\Omega)}^{2}
\leq 0,
\]
and hence
\[
\big\| (y - \Upsilon)^{+}(t) \big\|_{L^{2}(\Omega)}
\leq
\big\| (y - \Upsilon)^{+}(0) \big\|_{L^{2}(\Omega)} .
\]
Consequently,
\[
\big\| (y - \Upsilon)^{+}(t) \big\|_{L^{2}(\Omega)} = 0,
\quad \text{that is } y \leq \Upsilon .
\]}
Similarly, we have $y\geq -\Upsilon$.\\ Consequently, we  derive that
$$
|y(t, x)| \leq\left\|y_{0}\right\|_{L^{\infty}(\Omega)}-\gamma t, \quad \forall(t, x) \in Q_{0}.
$$
Then 
$y(t) \equiv 0$ pour tout  $t \geq\gamma^{-1}\left\|y_{0}\right\|_{L^{\infty}(\Omega)}$.

\emph{Case 2:}  if $y_{0}\in L^{2}(\Omega)$. Let $\theta>0$ and consider control (\ref{eq:control_lawy}).  
In this case the state of the system (\ref{systSvvb_simple})-(\ref{eq:control_lawy}) is given by:
\begin{equation}\label{kolchi}
y(t,.)= \begin{cases} S(t)y_{0}+ \int_{0}^{t} S(t-s)f(s)ds\  \text { if } 0\leq t\leq \theta, \\ y_{1}(t,.) \ \text { if}~ t>\theta.\end{cases}	
\end{equation}
Where $S(t)$ is the semigroup generated by $\Delta$ and $y_{1}$ is the solution of the following system:
\begin{eqnarray}\label{KAMAKkbbbbn}
	\begin{cases}
		\frac{\partial y_{1}}{\partial t}-\Delta y_{1}+ \rho~  \sign~  y_{1}-f\ni 0,~\text { in } (\theta,+\infty )\times \Omega,\\
		y_{1}=0, ~ \ \text { in}~  (\theta,+\infty ) \times \partial \Omega,\\
		y_{1}(\theta,.)=y(\theta),~ \ \text {in } ~ \hspace*{0.5cm} \Omega. 
	\end{cases}
\end{eqnarray}
Furthermore, according to (\cite{barbu1993analysis}, p. 31),  we have that 
$$
\left\|S(t)y_{0}\right\|_{L^{\infty}(\Omega)}\leq \hspace*{0.1cm} C_{1}\hspace*{0.1cm}t^{\frac{-n}{2}} \hspace*{0.1cm}\left\|y_{0}\right\|_{L^{2}(\Omega)}, \hspace*{0.1cm} 
$$ 
for all $ y_{0}\in L^{2}(\Omega)$ and all $t>0$. Here, the constant $C_{1}$ is independent of $y_{0}$. Hence,  we derive from {\color{blue} (\ref{kolchi})} that $y(\theta)\in L^{\infty}(\Omega)$. Then, through the same steps as in the first case, we conclude that  $y_{1}(t)=0$, for all $t\geq \theta+\gamma^{-1}\left\|y(\theta)\right\|_{L^{\infty}(\Omega)}$.
\end{proof}
\subsection{Finite-time stability results for nonlinear system \eqref{systSvvb}}
We  consider the following nonlinear feedback control:
 \begin{equation}\label{systSbnbbbbkamalfenzan}
u(y(t))= - \left\|B^{*}y(t)\right\|_{\mathcal{U}}^{-\mu}\hspace*{0.1cm} B^{*}y(t) \hspace*{0.1cm}\text{1}_{E} (y(t)),
\end{equation}
where  $\mu \in(0,\;1)$ and $E=\lbrace y\in \mathcal{H},\hspace*{0.1cm}  B^{*}y\neq 0\rbrace$. 

Substituting the control law from (\ref{systSbnbbbbkamalfenzan}) into system (\ref{systSvvb}), we obtain the following closed-loop system:
\begin{eqnarray}\label{systSresule}
	\begin{cases}
		\frac{d }{d t} y(t)-A y(t)+ F(y(t)) + G(y(t))=0, \quad    \hspace*{0.1cm} t>0,\\
		y(0)=y_{0},    \\
	\end{cases}
\end{eqnarray}
where  $$
G(y)= \begin{cases}  \left\|B^{*}y\right\|_{\mathcal{U}}^{-\mu}BB^{*}y, & B^{*}y \neq 0, \\ 0, & B^{*}y=0,\end{cases}
$$
and $F: \mathcal{H}\rightarrow \mathcal{H} $ is  defined by $F(y)=-f(y)y$.\\
We begin with the following result which   introduces the existence and uniqueness of the weak  solution of the
closed-loop system  (\ref{systSresule}).
\begin{theorem}\label{theo1}  Let $A$ generate a semigroup of contractions $S(t)$ on $\mathcal{H}$ and let   the assumption \text{$\text{(}\mathcal{\text{H}}_{\text{1}}\text{)}$} holds. Then,  for all $y_{0}\in \mathcal{H}$,  the system (\ref{systSresule}) has a unique weak solution $y\in C([0,+\infty), \mathcal{H})$.
\end{theorem}
\begin{proof}
To establish the well-posedness of  (\ref{systSresule}), we'll show that the operator $\mathcal{B}=-A+F_{1}+F_{2}+G+ \kappa I$ is maximal monotone, where $\kappa$ is the Lipschitz constant of $F_{2}$.\\
Since $A$   generates a   semigroup of contractions, the operator $-A$ is maximal monotone.  Furthermore, from the assumption \text{$\text{(}\mathcal{\text{H}}_{\text{1}}\text{)}$}, the operator $F_{1}+F_{2}+G+ \kappa I$ is  maximal monotone,  we deduce 
 that     $\mathcal{B}$ is  maximal monotone on $D(A)$.  According to (\cite{brezis1973operateurs}, p. 105),
  we  conclude that for all $y_{0} \in \mathcal{H}$,   the system  (\ref{systSresule}) has a unique weak solution $y\in C([0,+\infty), \mathcal{H})$.
\end{proof}

We can state our main result on finite-time stabilization of system (\ref{systSvvb}) as follows.
\begin{theorem}\label{theo2} Assume that the conditions of Theorem  \ref{theo1} are satisfied
and suppose  that \text{$\text{(}\mathcal{\text{H}}_{\text{2}}\text{)}$}  holds. Then,  the system  (\ref{systSvvb}) is globally finite-time stable under the following control:
\begin{equation}\label{con}
    u(y(t))= \begin{cases}  \left\|B^{*}y(t)\right\|_{\mathcal{U}}^{-\mu}B^{*}y(t), & y(t) \neq 0, \\ 0, & y(t)=0, \end{cases}
\end{equation}
where $\mu \in (0,1)$. Furthermore, the settling time admits the estimate  $T\leq\frac{\left\|y_{0}\right\|^{\mu}}{\beta^{1-\frac{\mu}{2}} \mu}$.
\end{theorem}
\begin{proof}
Under assumption \text{$\text{(}\mathcal{\text{H}}_{\text{2}}\text{)}$}, system (\ref{systSvvb}) with control (\ref{con})  is equivalent to system~\eqref{systSresule}. Therefore, by Theorem~\ref{theo1}, system~\eqref{systSresule} admits a unique weak solution for every initial datum \(y_0\in\mathcal H\). Hence, there is  a sequence $y_{0}^n$ in $D(A)$ such that $y_{0}^n \rightarrow y_{0}$, and the system  (\ref{systSresule}) with $y_{0}^n$ as initial state possesses a strong solution $y^n \in C([0,+\infty), \mathcal{H}$ verifying $y^n \rightarrow y$ uniformly on $[0, T^1]$ (for  any $T^1>0$ ) as $n \rightarrow \infty$.

 Then,  for a.e $t>0$,  we have $y^n(t) \in D(A)$  and   
$$ 
\begin{aligned}
\frac{\mathrm{d}}{\mathrm{d} t}y^n(t)-Ay^n(t)+ F(y^n(t)) +G(y^n(t)) =0,
\end{aligned}
$$
Now, in order to  prove the finite-time stability of the system   (\ref{systSresule}). We consider the Lyapunov function candidate  defined by $V(y)=\left\|y\right\|^{2} $, for all $y\in \mathcal{H}$. For a.e. $t >0$, the time derivative of $V$ along the trajectories  of the system  (\ref{systSresule}), yields
$$
\begin{aligned}
\frac{1}{2} \frac{d}{d t}\left\|y^n(t)\right\|^{2} &=\langle y^n(t), Ay^n(t)-F(y^n(t))-G(y^n(t))\rangle. \\
\end{aligned}
$$
By Assumption  \text{$\text{(}\mathcal{\text{H}}_{\text{2}}\text{)}$}, we find
$$
\begin{aligned}
\frac{1}{2} \frac{d}{d t}V(y^n(t))\leq 
\begin{cases}
		 -\beta^{1-\frac{\mu}{2}} V(y^n(t))^{1-\frac{\mu}{2}}, &  y^n(t) \neq 0, \\
		0, &  y^n(t)=0.
	\end{cases}
\\
\end{aligned}
$$
Hence,  according to the comparison principle, we have
$$ 
	\begin{cases}
	\begin{aligned}
		& V(y^n(t))^{\frac{\mu}{2}}\leq V(y_{0}^n)^{\frac{\mu}{2}} -  \beta^{1-\frac{\mu}{2}}\mu  t, & \text{ if }   t \leq \frac{V(y_{0}^n)^{\frac{\mu}{2}}}{ \beta^{1-\frac{\mu}{2}}\mu},\\
			& y^n(t)=0, & \text{ if }  t \geq  \frac{V(y_{0}^n)^{\frac{\mu}{2}}}{\beta^{1-\frac{\mu}{2}} \mu}.
		\end{aligned}
	\end{cases} 
$$
Then, letting  $n\rightarrow +\infty$,  we deduce that
$$ 
	\begin{cases}
	\begin{aligned}
		& \left\|y(t)\right\|^{\mu}\leq \left\|y_{0}\right\|^{\mu} -  \beta^{1-\frac{\mu}{2}}\mu  t,  \text{ if }   t \leq \frac{\left\|y_{0}\right\|^{\mu}}{ \beta^{1-\frac{\mu}{2}}\mu},\\
			& y(t)=0,  \text{ if }  t \geq  \frac{\left\|y_{0}\right\|^{\mu}}{\beta^{1-\frac{\mu}{2}} \mu}.
		\end{aligned}
	\end{cases} 
$$
This achieves the proof.
\end{proof}
\begin{remark}
     The proposed control law in \eqref{systSbnbbbbkamalfenzan} is nonsingular because \(\lim_{t \to T} y(t) = 0\) and \(\lim_{y(t) \to 0} \|u(y(t))\| = 0\).
   \end{remark}

\subsection{Application: Heat equation}

Let $\Omega=(0,1)$  and let us consider the following   multi-dimensional heat equation:
\begin{IEEEeqnarray}{c}\label{exemple}
\left\{
\begin{aligned}
&y_t(x, t) = \Delta y(x, t) + f(y(x, t)) y(x, t) \\
& + \sqrt{a(x)} \, u(y(x, t)), \quad (x, t) \in \Omega \times (0,+\infty) \\
&y(x, t) = 0, \quad (x, t) \in \partial \Omega \times (0,+\infty) \\
&y(x, 0) = y_0, \quad x \in \Omega
\end{aligned}
\right.
\end{IEEEeqnarray}
where  $y (t)=y (\cdot,t) $ is the state, $a \in L^{\infty}(\Omega)$ and $a(x) \geq c$ a.e. $x \in \Omega$ for some $c>0$.\\
We can write  (\ref{exemple}) in the form of  (\ref{systSvvb}) on $L^{2}(\Omega )$  if we set $Ay=\Delta y$ for $y\in D(A)=\mathcal{H}_{0}^{1}(\Omega ) \cap \mathcal{H}^2(\Omega )$,  the spectrum of $A$ is given by the
simple eigenvalues $\lambda_{j}= -(j\pi)^{2}$ and eigenfunctions $\varphi_{j}(x)=\sqrt 2 $ $ sin(j\pi x)$, $j\geq 1$. Furthermore $A$ generates a contraction semigroup given by $S(t) y=\sum\limits_{j=1}^{\infty} \exp \left(\lambda_{j} t\right)\langle y, \phi_j\rangle \phi_j$, for all $y\in L^{2}(\Omega )$ and all $t\geq 0$. \\
Consider the bounded operator $B :L^{2}(\Omega) \rightarrow L^{2}(\Omega)$ defined by $Bz= \sqrt{a(.)}\hspace*{0.1cm}z$. Then, the assumption \text{$\text{(}\mathcal{\text{H}}_{\text{2}}\text{)}$} is verified with $\beta=c$.  Let us consider the following control: 
\begin{gather}\label{exemple11}
u(y (\cdot,t)) = \begin{cases} 
\begin{aligned}[t]
&-\|\sqrt{a(\cdot)} y (\cdot,t)\|^{- \mu} \sqrt{a(\cdot)}\\
&\quad \times y (\cdot,t)
\end{aligned} & \text{if } y (\cdot,t) \neq 0, \\ 
0 & \text{otherwise.}
\end{cases}
\end{gather}
where  $\mu \in(0,\;1)$. This leads to the closed-loop system:
\begin{IEEEeqnarray}{c}\label{exemple1}
\left\{
\begin{aligned}
y_t(x, t) = \Delta y(x, t) + f(y(x, t)) y(x, t) \\
 - \|\sqrt{a(x)} y(x,t)\|^{-\mu} a(x) y(x, t), \\
 (x, t) \in \Omega \times (0,+\infty) \\
y(x, t) = 0, \quad (x, t) \in \partial \Omega \times (0,+\infty) \\
y(x, 0) = y_0, \quad x \in \Omega
\end{aligned}
\right.
\end{IEEEeqnarray}
The problem of distributed finite-time control for the  heat system (\ref{exemple1})  with $f=0$ has been extensively studied in  \cite{sob}, while the specific case where $a(x)=1$ has been considered in  \cite{2333}. In this work, we extend the analysis by considering a more general function $f$, defined as  $f(y)= -\frac{\|\sqrt{a(\cdot)} \hspace*{0.1cm}y\|^{- \mu} \hspace*{0.1cm}a(\cdot)}{1+\|\sqrt{a(\cdot)} \hspace*{0.1cm}y \|^{2 } }$.  
Using this formulation, the assumption \text{$\text{(}\mathcal{\text{H}}_{\text{1}}\text{)}$} is naturally satisfied, ensuring the well-posedness of the problem under our proposed framework.

Consequently,  applying Theorem \ref{theo2}, the feedback control (\ref{exemple11})  guarantees the global finite-time stability
of the system (\ref{exemple1}).
\subsection{Simulations}
In this section, we present numerical simulations to illustrate the theoretical results obtained in the previous sections. 
The evolution of the state of the system described by equation (\ref{exemple1}) for the case where  $a(x)=x^{2 }+0.01$ with the initial condition $y_0 (x)=5x(1-x)$, for all $x\in \Omega=(0,1)$,  is illustrated in  the following figure:
\begin{figure}[h!]
	\centerline{\includegraphics[scale=0.7,width=90mm]{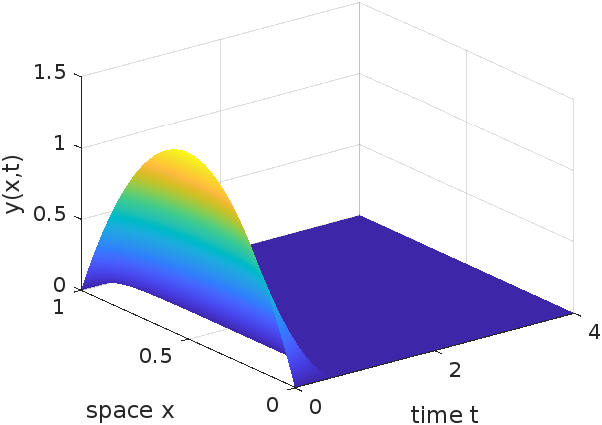}}
    	\caption{Evolution of  the state  $y(x,t)$   } 
	\end{figure}
    The figure below represents the evolution of the norm of the state $y$ for different values of the control parameter $\mu$. From the plot, it is evident that the system stabilizes faster for higher values of $\mu$. Specifically,  feedback control (\ref{exemple11}) with larger $\mu=0.8$  brings the system to equilibrium more quickly than control  with smaller $\mu=0.2$. This behavior highlights the influence of 
$\mu$ on the stabilization process: as $\mu$ increases, the control action becomes more aggressive, leading to faster stabilization of the system.

	\begin{figure}[h!]
		\centerline{\includegraphics[scale=0.6,width=90mm]{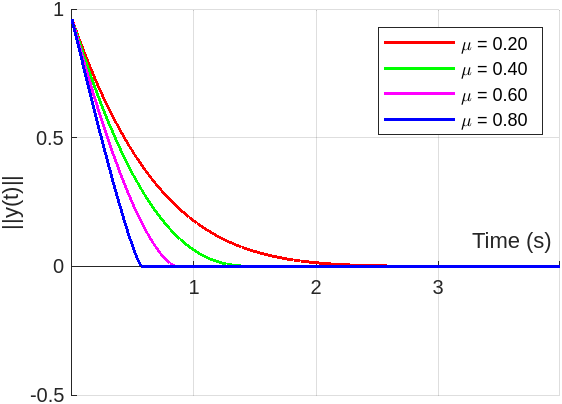}}
        \caption{The evolution of $\|y(t)\|$ for different values  of $\mu$ }
	\end{figure}

\section{Conclusions}
This paper has addressed the finite-time stabilization of a class of nonlinear infinite-dimensional systems. We first considered a bounded matched perturbation in its linear form and demonstrated that using a set-valued function achieves both the convergence objective and perturbation rejection. We then extended our analysis to a class of nonlinear systems, designing a feedback control law that ensures finite-time stability of the closed-loop system. The theoretical results were established using Lyapunov theory for convergence analysis and maximal monotone theory for proving well-posedness and the existence of solutions. To validate the effectiveness of the proposed approach, a heat equation was examined as an application of the main results.

Future research will focus on extending these results to more complex partial differential equations, particularly those arising in fluid dynamics and distributed parameter systems. Additionally, we aim to investigate the impact of perturbations in boundary control.

\bibliographystyle{plain}  
\balance
\bibliography{Bib}  

\end{document}